%% file: gamma-en.tex
\documentclass[11pt,a4paper]{article}

\usepackage[cp1251]{inputenc}
\usepackage[english]{babel}

\newtheorem{theorem}{Theorem}[section]
{\bf}{\it }
\newtheorem{corollary}{Corollary}[section]

\title{Searching isomorphic graphs}

\author{Anatoly D. Plotnikov\footnote{Corresponding author
E-mail address: a.plotnikov@list.ru}
}

\date{East Ukrainian National University,\\
Severodonetsk,  Ukraine}

\begin{document}
\maketitle

\begin{abstract}
To determine that two given undirected graphs are isomorphic, we construct for them auxiliary graphs, using the 
breadth-first search. This makes capability to position vertices in each digraph with respect to
each other. If the given graphs are isomorphic, in each of them we can find such positionally equivalent auxiliary 
digraphs that have the same mutual positioning of vertices. Obviously, if the given graphs are isomorphic, then 
such equivalent digraphs exist. Proceeding from the arrangement of vertices in one of the digraphs, we try to
determine the corresponding vertices in another digraph. As a result we develop the algorithm for constructing 
a bijective mapping between vertices of the given graphs if they are isomorphic. The running time of the 
algorithm equal to $O(n^5)$, where $n$ is the number of graph vertices.
\end{abstract}

\vspace{1pc}
{\bf MCS2000:} 05C85, 68Q17.

{\bf Key words:} graph, isomorphism, bijective mapping, isomorphic graphs, algorithm, graph isomorphism problem.

\section{Introduction}

Let $L_n$ is the set of all $n$-vertex undirected graphs without loops and multiple edges.

Let, further, there is a graph $G=(V_G,E_G)\in L_n$, where $V_G=\{v_1,v_2,\ldots,$ $v_n\}$ is the set of vertices 
and $E_G=\{e_1,e_2,\ldots,e_m\}$ is the set of edges of the graph $G$. Local degree $deg(v)$ of the vertex $v\in V_G$ 
is the number of edges, that is incident to the vertex $v$. Every graph $G\in L_n$ can be characterized by the vector 
$D_G=(deg(v_{i_1}),deg(v_{i_2}),\ldots,deg(v_{i_n}))$ of the local vertex degrees, where $deg(v_i)\leq deg(v_j)$ if $i<j$.

Graphs $G=(V_G,E_G)$, $H=(V_H,E_H)\in L_n$ are called isomorphic if between their vertices there exists one-to-one 
(bijective) mapping $\varphi:$ $V_G\leftrightarrow V_H$ such that if $e_G=\{v,u\}\in E_G$ then the corresponding 
edge is $e_H=\{\varphi(v)\varphi(u)\}\in E_H$, and conversely \cite{west}. We say that the mapping $\varphi$ converts 
the graph $G$ into the graph $H$ and conversely.

The problem of determining the isomorphism of two given undirected graphs is used to solve chemical problems, and 
to optimize programs \cite{varmuza} --- \cite{wale} and others. Effective (polynomial-time) algorithms for solving 
this problem were found for some narrow classes of graphs \cite{aho,garey}. However for the general case, effective 
methods for determining the isomorphism of graphs are not known \cite{babai}.

The purpose of this article is to propose the polynomial-time algorithm for  searching  isomorphic graphs.

\section{Basic definitions}

Let there be a graph $G\in L_n$.

Choose some vertex $v\in V_G$. The set of all vertices of the graph $G$, adjacent to vertex $v$, we call the neighborhood of this vertex 
of the 1st level.

Suppose that we have constructed the neighborhood of the vertex $v$ of $(k-1)$-th level. Then the set of all vertices adjacent to at 
least one vertex $(k-1)$-th level, we call the neighborhood of the vertex $k$-th level $(0\leq k\leq n-1)$. Such neighborhood we denote 
$N_G^{(k)}(v)$. For convenience, we assume that the vertex $v$ forms the neighborhood of the zero level.

Found neighborhoods allow us to construct for the vertex $v$ of the auxiliary directed graph $\vec G(v)$ in the following way.

Each neighborhood $N_G^{(k)}(v)$ of the graph $G$ forms $k$-th line of the digraph. If the edge of the graph $G$ connects the vertex $v_1$ 
of $k_1$-th level with the vertex $v_2$ of $k_2$ level, and $k_1<k_2$, then this edge is replaced by the arc $(v_1,v_2)$. If $k_1=k_2$, then 
this edge is replaced by two arcs $(v_1,v_2)$ and $(v_2,v_1)$. We say that the constructed digraph is induced by the vertex $v$ of the graph $G$.

\begin{figure}[ht]
\centering
\mbox{\input{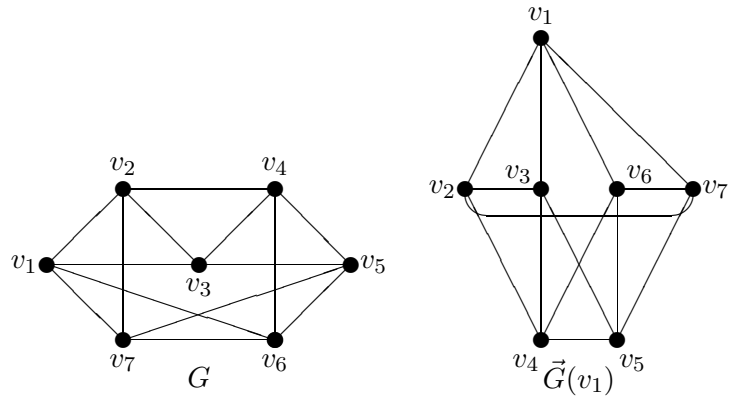}}
\caption{The graph $G$ and the auxiliary digraph $\vec G(v_1)$}
\label{gg1}
\end{figure}

Fig. \ref{gg1} shows the graph $G$ and the auxiliary digraph $\vec G(v_1)$, induced by the vertex $v_1$.

Each vertex of an auxiliary graph will be characterized by two vectors.

The input characteristic of the vertex $v$ of the digraph $\vec G(v)$ is called the 
vector $I_{v}=(i_1,\ldots,i_q)$, where the elements $i_1,\ldots,i_q$ are the line 
numbers of the digraph, written in order of increase. These numbers indicate the 
lines, from which arcs come into the vertex $v$. If into the vertex $v$ several 
arcs come from the same line, the line number is recorded in the vector $I_v$ the 
corresponding number of times.

The output characteristic of a vertex $v$ of the digraph $\vec G(v)$ is called the 
vector $O_{v}=(i_1,\ldots,i_q)$, where the elements $i_1,\ldots,i_q$ are the 
line numbers of the digraph, also written in order of increase. These 
numbers indicate those lines, in which arcs come from vertex $v$. If from the 
vertex $v$ several arcs come into vertices of the same line, then the line number are 
recorded in the vector $I_v$ the corresponding number of times.

Characteristic of vertex $v$ of the digraph $\vec G(v)$ will be called the input and output characteristics of this vertex. The characteristics of the two vertices $v_1$, $v_2$ are equal if them the input and output characteristics are equal, respectively.

Find the vertex characteristics of the digraph $\vec G(v_1)$, is shown in Fig. \ref{gg1}. The results are presented in 
the table \ref{tb1}.
 
\medskip

\begin{table}[ht]
\centering
\caption{The vertex characteristics of the digraph $\vec G(v_1)$}
\label{tb1}
\begin{tabular}{|l|l|l|l|}
\hline
$I_{v_1}=\oslash$; & $I_{v_2}=(0,1,1)$; & $I_{v_3}=(0,1)$; & $I_{V_4}=(1,1,1,2)$;\\
$O_{v_1}=(1,1,1,1)$; & $O_{v_2}=(1,1,2)$; & $O_{v_3}=(1,2,2)$; & $O_{v_4}=(2)$;\\
\hline
$I_{v_5}=(1,1,1,2)$; & $I_{v_6}=(0,1)$; & $I_{v_7}=(0,1,1)$; & \\
$O_{v_5}=(2)$; & $O_{v_6}=(1,2,2)$; & $O_{v_7}=(1,1,2)$. & \\
\hline
\end{tabular}
\end{table}
\medskip

Let there are auxiliary directed graphs $\vec G(v_1)$ and $\vec G(v_2)$, induced by the vertices $v_1$ and $v_2$, possibly belonging to different graphs. The 
directed graphs $\vec G(v_1)$ and $\vec G(v_2)$ are called positionally equivalent if the lines of digraphs of the same level contain the same number of 
vertices having respectively equal (input and output) characteristics.

A vertex $x\in V_G$ will be called unique if the digraph $\vec G(v)$ does not exist another vertex with characteristics equal to the characteristics of the vertex $x$. 

Note that the vertex $v\in V_G$, that induces the auxiliary digraph $\vec G(v)$, is always unique in this digraph.

\section{The basics of the algorithm}

Next, we will consider pairs of graphs $G,H\in L_n$, having equal number of vertices $n$, equal number of edges $m$ and equal vectors of the local degrees $D_G=D_H$. It needs to determine the isomorphism of the given graphs and, if they are isomorphic, then find the bijective mapping $\varphi$ between their vertices.

The idea of finding bijective mapping $\varphi$ between the vertices of the vertex set of graphs $G,H\in L_n$ is the following. We assume that the graphs $G$ and $H$ are isomorphic. Naturally that the required mapping can only be found when graphs $G$ and $H$ is indeed isomorphic. If in the process of finding the mapping $\varphi$ cannot be found, then the given graphs are not isomorphic.

\begin{theorem}
\label{thm1}
Let the graphs $G$ and $H$ are isomorphic. Then there exist at least two vertices $v\in V_G$ and $u\in V_H$ such that induce two auxiliary positionally equivalent digraphs $\vec G(v)$ and $\vec H(u)$.

\end{theorem}

{\bf Proof}. The construction of the auxiliary directed graphs depends only on the location of graph vertices relative to the neighborhood of the induced vertex $v$ or $u$ and does not depend on the vertex names of the graphs $G$ and $H$. Therefore, because graphs $G$ and $H$ are isomorphic and we have the identical procedure of constructing the auxiliary digraphs, we will be found the auxiliary positionally equivalent digraphs $\vec G(v)$ and $\vec H(u)$.$\diamondsuit$

\begin{theorem}
\label{thm2}
If graphs $G$ and $H$ are isomorphic and two auxiliary positionally 
equivalent digraphs $\vec G(v)$ and $\vec H(u)$ are found, then any bijective mapping $\varphi$, which convert the graph $G$ into the graph $H$ (and conversely), is determined by pairs of vertices of the digraphs with equal characteristics.
\end{theorem}

{\bf Proof}. The assertion of the Theorem \ref{thm2} is true as if to assume  contrary, we will get that the mapping $\varphi$ converts one vertex to another with different characteristics. This is contrary to the concept of isomorphism of graphs.$\diamondsuit$

\begin{corollary}
\label{cor1}
Let the graphs $G$ and $H$ are isomorphic and two auxiliary positionally equivalent digraphs $\vec G(v)$ and $\vec H(u)$ are constructed. Let, further, it was found $t$  unique vertices  in these digraphs, the corresponding pairs of which have equal characteristics. Then, the pairs of these vertices belong to the mapping $\varphi$.
\end{corollary}

\medskip

It is easy to understand that the search of vertex pairs that belong to the binary mapping $\varphi$, the equality of the vertex characteristics in the auxiliary digraphs are not sufficient if these vertices have incoming and/or outgoing arcs, connecting vertices of the same line of the auxiliary digraph. 

In the isomorphic graphs $G$ and $H$, we can find vertices $v\in V_G$ and $u\in V_H$ such that induce auxiliary positionally equivalent digraphs $\vec G(v)$ and $\vec H(u)$ respectively. In accordance with the Theorem \ref{thm1}, these vertices always exist.

The desired bijective mapping $\varphi$ can be represented as the perfect matching in the bipartite graph, induced by the vertices of the auxiliary positionally equivalent digraphs $\vec G(v)$ and $\vec H(u)$. This bipartite graph we will call the virtual. In this bipartite graph, any vertex $x\in V_G$ is connected by the virtual edge with all vertices of the digraph $\vec H(u)$, which have same characteristics as $x$. It is clear that the bijective mapping will correspond to virtual perfect matching in the bipartite graph. Unfortunately, it is not every perfect matching corresponds to the bijective mapping $\varphi$.

\begin{figure}[ht]
\centering
\mbox{\input{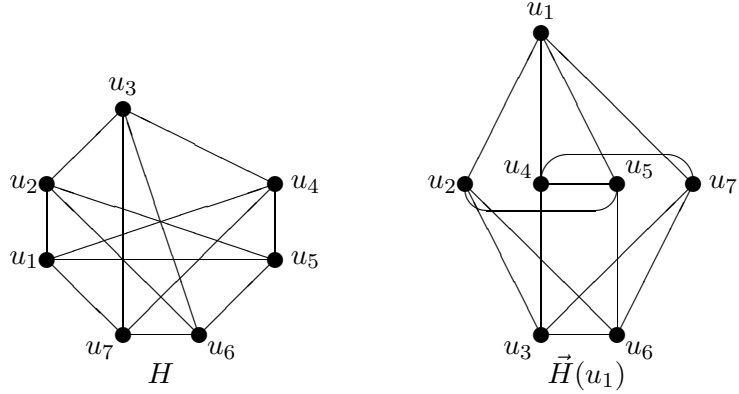}}
\caption{The graph $H$ and the auxiliary digraph $\vec H(u_1)$.}
\label{gg2}
\end{figure}

Fig. \ref{gg2} shows the graph $H$ and the auxiliary digraph $\vec H(u_1)$, induced by the vertex $u_1$.

We  find the vertex characteristics of the auxiliary digraph $\vec H(u_1)$. The results shown in the table \ref{tb2}.

\medskip

\begin{table}[ht]
\centering
\caption{The vertex characteristics of the digraph $\vec H(u_1)$}
\label{tb2}
\begin{tabular}{|l|l|l|l|}
\hline
$I_{u_1}=\oslash$; & $I_{u_2}=(0,1)$; & $I_{u_3}=(1,1,1,2)$; & $I_{u_4}=(0,1,1)$;\\
$O_{u_1}=(1,1,1,1)$; & $O_{u_2}=(1,2,2)$; & $O_{u_3}=(2)$; & $O_{u_4}=(1,1,2)$;\\
\hline
$I_{u_5}=(0,1,1)$; & $I_{u_6}=(1,1,1,2)$; & $I_{u_7}=(0,1)$; & \\
$O_{u_5}=(1,1,2)$; & $O_{u_6}=(2)$; & $O_{u_7}=(1,2,2)$. & \\
\hline
\end{tabular}
\end{table}
\medskip

It is easy to see that the auxiliary digraphs $\vec G(v_1)$ (see Fig. \ref{gg1}) and $\vec H(u_1)$ are positionally equivalent. In these digraphs, there are only two unique vertices $v_1$ and $u_1$. Other vertices with equal characteristics form the virtual bipartite graphs, induced by the following vertex pairs: $(\{v_3,v_6\},\{u_2,u_7\})$, $\{v_4,v_5\},\{u_3,u_6\})$ and $(\{v_2,v_7\},\{u_4,u_5\})$.

\begin{figure}[htbp]
\centering
\mbox{\input{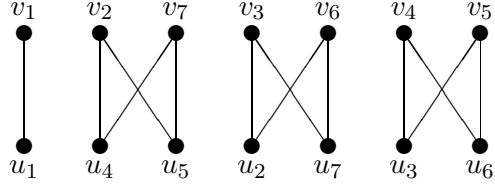}}
\caption{The virtual bipartite graph.}
\label{gg3}
\end{figure}

Fig. \ref{gg3} shows the virtual bipartite graph for the auxiliary positionally equivalent digraphs $\vec G(v_1)$ and $\vec H(u_1)$.

In this case, we have one unique vertex in each of the digraphs $\vec G(v)$ and $\vec H(u)$. Therefore, the pair of vertices $\{v,u\}$ belongs to the desired mapping $\varphi$. Remove these vertices from the corresponding graphs. We will get graphs $G_1$ and $H_1$.

\begin{figure}[ht]
\centering
\mbox{\input{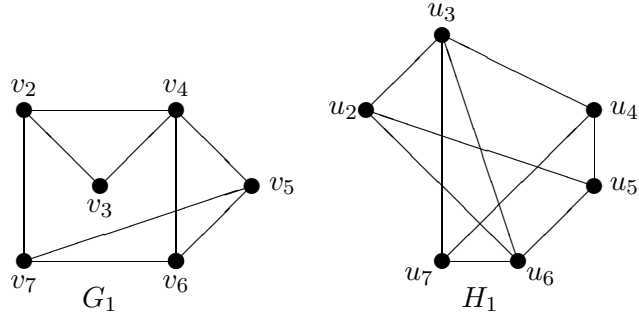}}
\caption{The graphs $G_1$ and $H_1$.}
\label{gg4}
\end{figure}

Fig. \ref{gg4} shows graphs $G_1$ and $H_1$.

Thus, we again have the problem of finding the bijective mapping $\varphi_1\subset \varphi$, which converts the graph $G_1$ to the graph $H_1$ and conversely.

\begin{figure}[ht]
\centering
\mbox{\input{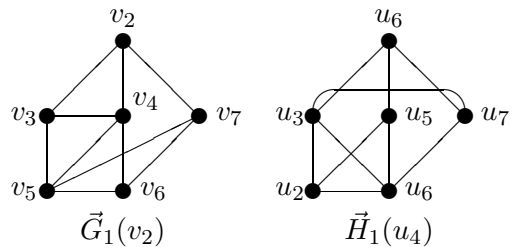}}
\caption{The digraphs $\vec G_1(v_2)$ and $\vec H_1(u_4)$.}
\label{gg5}
\end{figure}

In the graph $G_1$ we choose the vertex $v_2$, and construct the corresponding auxiliary digraph (see Fig. \ref{gg5}). 
We find the vertex characteristics of the digraph $\vec G_1(v_2)$. The result of the calculation is placed in the 
table \ref{tb3}.

\medskip

\begin{table}[ht]
\centering
\caption{The vertex characteristics of the digraph $\vec G_1(v_2)$}
\label{tb3}
\begin{tabular}{|l|l|l|}
\hline
$I_{v_2}=\oslash$; & $I_{v_3}=(0,1)$; & $I_{v_4}=(0,1)$; \\
$O_{v_2}=(1,1,1)$; & $O_{v_3}=(1,2)$; & $O_{v_4}=(1,2,2)$; \\
\hline
$I_{v_5}=(1,1,1,2)$; & $I_{v_6}=(1,1,2)$; & $I_{v_7}=(0)$; \\
$O_{v_5}=(2)$; & $O_{v_6}=(2)$; &  $O_{v_7}=(2,2)$;  \\
\hline
\end{tabular}
\end{table}
\medskip

In the graph of $H_1$ we choose the vertex $u_4$, and construct the corresponding auxiliary graph (see Fig. \ref{gg5}). 
Find the vertex characteristics of the digraph $\vec H_1(u_4)$. The result of the calculation is placed in the 
table \ref{tb4}.

\medskip

\begin{table}[ht]
\centering
\caption{The vertex characteristics of the digraph $\vec H_1(u_4)$}
\label{tb4}
\begin{tabular}{|l|l|l|}
\hline
$I_{u_4}=\oslash$; & $I_{u_3}=(0,1)$; & $I_{u_5}=(0)$; \\
$O_{u_4}=(1,1,1)$; & $O_{u_3}=(1,2,2)$; & $O_{u_5}=(2,2)$; \\
\hline
$I_{u_7}=(0,1)$; & $I_{u_2}=(1,1,2)$; & $I_{u_6}=(1,1,1,2)$; \\
$O_{u_7}=(1,2)$; & $O_{u_2}=(2)$; &  $O_{u_6}=(2)$;  \\
\hline
\end{tabular}
\end{table}
\medskip

Comparing the vertex characteristics of auxiliary digraphs, we see that the digraphs $\vec G_1(v_2)$ and $\vec H_1(u_4)$ is the positional equivalent. Moreover, in each of the constructed digraph, each vertex is unique.

At once, we can construct the mapping $\varphi_1$, choosing vertex pairs of the digraphs $\vec G_1(v_2)$ and $\vec H_1(u_4)$ with equal characteristics. Adding to $\varphi_1$ of the previously found pair $(v_1,u_1)$, we get the required mapping 
\[\varphi=\{(v_1,u_1), (v_2,u_4),(v_3,u_7),(v_4,u_3),(v_5,u_6),(v_6,u_2),(v_7,u_5)\}.
\]
To perform the verification of the obtained result, consider the matching  edges of graphs $G$ and $H$ defined found the bijective mapping.

\medskip

\begin{tabular}{cc}
\begin{tabular}{c}
$\{v_1,v_2\} \leftrightarrow \{u_1,u_4\}$ \\
$\{v_1,v_3\} \leftrightarrow \{u_1,u_7\}$ \\
$\{v_1,v_6\} \leftrightarrow \{u_1,u_2\}$ \\
$\{v_1,v_7\} \leftrightarrow \{u_1,u_5\}$ \\
$\{v_2,v_3\} \leftrightarrow \{u_4,u_7\}$ \\
$\{v_2,v_4\} \leftrightarrow \{u_4,u_3\}$ \\
$\{v_2,v_7\} \leftrightarrow \{u_4,u_5\}$\\ 
\end{tabular}
&
\begin{tabular}{c}
$\{v_3,v_4\} \leftrightarrow \{u_7,u_3\}$ \\
$\{v_3,v_5\} \leftrightarrow \{u_7,u_6\}$ \\
$\{v_4,v_5\} \leftrightarrow \{u_3,u_6\}$ \\
$\{v_4,v_6\} \leftrightarrow \{u_3,u_2\}$ \\
$\{v_5,v_6\} \leftrightarrow \{u_6,u_2\}$ \\
$\{v_5,v_7\} \leftrightarrow \{u_6,u_5\}$ \\
$\{v_6,v_7\} \leftrightarrow \{u_2,u_5\}$\\ 
\end{tabular}\\
\end{tabular}

All matches are correct.

\section{Vertex characteristics of the digraphs}

The example above illustrates our approach to solving problem of finding the bijective mapping $\varphi$ between the vertices in isomorphic graphs $G$ and $H$.

The essence of this approach consists in the following steps.

\begin{itemize}
\item In the graph $G$, choose the vertex $v$.
\item Construct the auxiliary digraph $\vec G(v)$.
\item In the graph $H$, choose the vertex $u$ such that the auxiliary digraph $\vec H(u)$, which is positionally equivalent to the digraph $\vec G(v)$. If such vertex is not found, then terminate computation as graphs $G$ and $H$ are not 
isomorphic.
\item In the positionally equivalent digraphs $\vec G(v)$ and $\vec H(u)$, 
find all unique vertices. Form vertex pairs $(v,u)$, having equal 
characteristics, and record them. Delete all unique vertices from the graphs $G$ and $H$.
\item If the graphs are obtained after the removal of unique vertices is not 
empty, then repeat the above procedure again. Otherwise, recorded vertex pairs 
form the desired bijective mapping $\varphi$.
\end{itemize}

The conception of positional equivalence of auxiliary digraphs requires clarification in the conditions when the algorithm at each iteration is dealing with the changed set of vertices of the graph. This concept is due to the vertex characteristics of the considered digraphs. In the process of work of the search algorithm of the bijective mapping, we have to take into account the history of the choice of vertices for mapping $\varphi$ in the previous iteration of the algorithm.

Let us explain the above by the example.

\begin{figure}[ht]
\centering
\mbox{\input{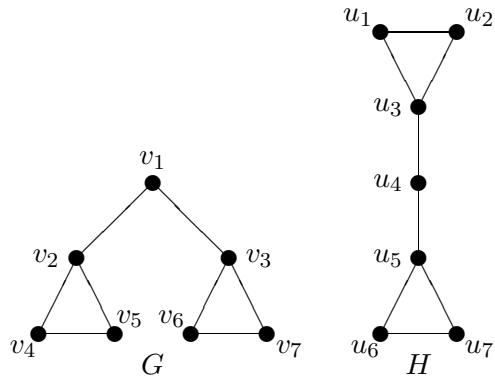}}
\caption{The graphs $G$ and $H$.}
\label{b1}
\end{figure}

Suppose there are two graphs $G$ and $H$ (see Fig. \ref{b1}), bijective mapping  the vertices of which we want to find.

In the graph $G$, we choose the vertex $v_1$ and construct the corresponding auxiliary graph $\vec G(v_1)$. On form, 
it coincides with the figure of the graph $G$. Therefore, we immediately find 
the vertex characteristics of the digraph $\vec G(v_1)$.
 
The results are presented in the table \ref{tb5}.
\medskip

\begin{table}[ht]
\centering
\caption{The vertex characteristics of the digraph $\vec G(v_1)$}
\label{tb5}
\begin{tabular}{|l|l|l|l|}
\hline
$I_{v_1}=\oslash$; & $I_{v_2}=(0)$; & $I_{v_3}=(0)$; & $I_{V_4}=(1,2)$;\\
$O_{v_1}=(1,1)$; & $O_{v_2}=(2,2)$; & $O_{v_3}=(2,2)$; & $O_{v_4}=(2)$;\\
\hline
$I_{v_5}=(1,2)$; & $I_{v_6}=(1,2)$; & $I_{v_7}=(1,2)$; & \\
$O_{v_5}=(2)$; & $O_{v_6}=(2)$; & $O_{v_7}=(2)$. & \\
\hline
\end{tabular}
\end{table}
\medskip

Further, when we construct the auxiliary directed graph for the vertices of the graph $H$, we find the digraph 
$\vec H(u_4)$, induced by the vertex $u_4$. It is easy to see that it will coincide with the figure of 
the digraph $\vec G(v_1)$. Find the vertex characteristics of this digraph.

The results are presented in the table \ref{tb6}.
\medskip

\begin{table}[ht]
\centering
\caption{The vertex characteristics of the digraph $\vec H(u_4)$}
\label{tb6}
\begin{tabular}{|l|l|l|l|}
\hline
$I_{u_1}=(1,2)$; & $I_{u_2}=(1,2)$; & $I_{u_3}=(0)$; & $I_{u_4}=\oslash$;\\
$O_{u_1}=(2)$; & $O_{u_2}=(2)$; & $O_{u_3}=(2,2)$; & $O_{u_4}=(1,1)$;\\
\hline
$I_{u_5}=(0)$; & $I_{u_6}=(1,2)$; & $I_{u_7}=(1,2)$; & \\
$O_{u_5}=(2,2)$; & $O_{u_6}=(2)$; & $O_{u_7}=(2)$. & \\
\hline
\end{tabular}
\end{table}
\medskip

It is easy to see that the auxiliary digraphs $\vec G(v_1)$ and $\vec H(u_4)$ positionally equivalent. They have by 
the single unique vertex of $v_1$ and $u_4$, respectively.

\begin{figure}[ht]
\centering
\mbox{\input{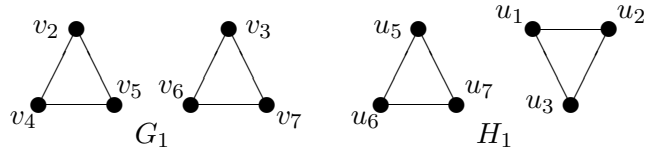}}
\caption{Graphs $G_1$ and $H_1$.}
\label{b2}
\end{figure}

Remove the vertices $v_1$ and $u_4$ of graphs $G$ and $H$. We obtain disconnected graphs $G_1$ and $H_1$ is 
shown in Fig. \ref{b2}.

In the subgraph of $G_1$, we choose the vertex $v_2$ and construct the auxiliary digraph $\vec G_1(v_2)$. His form is 
the same as the first connected component of the graph. We find vertex characteristics of the digraph $\vec G_1(v_2)$.

The results are presented in the table \ref{tb7}.
\medskip

\begin{table}[ht]
\centering
\caption{The vertex characteristics of the digraph $\vec G_1(v_2)$}
\label{tb7}
\begin{tabular}{|l|l|l|}
\hline
$I_{v_2}=\oslash$; & $I_{v_4}=(0,1)$; & $I_{v_5}=(0,1)$;\\
$O_{v_2}=(1,1)$; & $O_{v_4}=(1)$; & $O_{v_5}=(1)$.\\
\hline
\end{tabular}
\end{table}
\medskip

In the graph $H_1$, we choose the vertex $u_1$ and construct the auxiliary digraph $\vec H_1(u_1)$. It is easy 
to see that its form will be coincide with the digraph $\vec G_1(v_2)$. We find vertex characteristics of the new 
digraph.

The results are presented in the table \ref{tb8}.
\medskip

\begin{table}[ht]
\centering
\caption{The vertex characteristics of the digraph $\vec H_1(u_1)$}
\label{tb8}
\begin{tabular}{|l|l|l|}
\hline
$I_{u_1}=\oslash$; & $I_{u_2}=(0,1)$; & $I_{u_3}=(0,1)$;\\
$O_{u_1}=(1,1)$; & $O_{u_2}=(1)$; & $O_{u_3}=(1)$.\\
\hline
\end{tabular}
\end{table}
\medskip

The resulting digraphs are positionally equivalent. Here vertices $v_2$ and $u_1$  are the unique, having equal 
characteristics. However, these pair of vertices $(v_2,u_1)$ will not belong to the bijective mapping $\varphi$ 
of graphs $G$ and $H$.

It happened because when considering graphs $G_1$ and $H_1$ we did not take into account the previous stage 
of calculations, when it was found the couple of unique vertices $v_1$, $u_4$, remote subsequently of the graphs 
$G$ and $H$.
\medskip

Each vertex of the auxiliary graph let's characterise by the old and new input and output characteristics in the 
following way.

Let in the given graphs $G$ and $H$,  positionally equivalent digraphs $\vec G(v)$ and $\vec H(u)$ were found. 
The calculated characteristics of the vertices of the digraphs, we fix for graph vertices $G$ and $H$. Subsequently, 
after removing the unique vertices of the graphs, the vertex characteristics for new digraphs are finding and they 
join previously found and fixed characteristics.

For definiteness, we assume that the new vertex characteristics of the digraphs are always located on the first 
``floor'' of the building from the vectors of input and output characteristics. The previously found characteristics 
moved on one ``floor'' up.

We assume vertex characteristics of the digraphs equal if their vector characteristics are equal on the respective 
``floors''. Similarly, two of the digraph call positionally equivalent if the line digraphs of the same level contain 
the same number of vertices having respectively equal to (input and output) characteristics.

We will write characteristics of vertices of the digraphs $\vec G_1(v_2)$ and $\vec H_1(u_1)$, obtained earlier.

We have the following characteristics for the digraphs $\vec G_1(v_2)$.
\medskip

\begin{table}[ht]
\centering
\caption{The vertex characteristics of the digraphs $\vec G_1(v_2)$}
\label{tb9}
\begin{tabular}{|l|l|l|}
\hline
$I_{v_2}=(0)$; & $I_{V_4}=(1,2)$; & $I_{v_5}=(1,2)$;\\
$I^{\prime}_{v_2}=\oslash$; & $I^{\prime}_{v_4}=(0,1)$; & $I^{\prime}_{v_5}=(0,1)$;\\
$O_{v_2}=(2,2)$; &  $O_{v_4}=(2)$; & $O_{v_5}=(2)$;\\
$O^{\prime}_{v_2}=(1,1)$; & $O^{\prime}_{v_4}=(1)$; & $O^{\prime}_{v_5}=(1)$.\\
\hline
\end{tabular}
\end{table}
\medskip

For the digraph $\vec H_1(u_1)$, we have the following characteristics (see table \ref{tb10}).
\medskip

\begin{table}[ht]
\centering
\caption{The vertex characteristics of the digraph $\vec H_1(u_1)$}
\label{tb10}
\begin{tabular}{|l|l|l|}
\hline
$I_{u_1}=(1,2)$; & $I_{u_2}=(1,2)$; & $I_{u_3}=(0)$;\\
$I^{\prime}_{u_1}=\oslash$; & $I^{\prime}_{u_2}=(0,1)$; & $I^{\prime}_{u_3}=(0,1)$;\\
$O_{u_1}=(2)$; & $O_{u_2}=(2)$; & $O_{u_3}=(2,2)$;\\ 
$O^{\prime}_{u_1}=(1,1)$; & $O^{\prime}_{u_2}=(1)$; & $O^{\prime}_{u_3}=(1)$.\\
\hline
\end{tabular}
\end{table}
\medskip

It is easy to see that the constructed digraphs are not positionally equivalent.

In the graph of $H_1$, we choose the vertex $u_3$, and construct the auxiliary digraph $\vec H_1(u_3)$ and find 
the vertex characteristics of the digraph.

The results are presented in the table \ref{tb11}.
\medskip

\begin{table}[ht]
\centering
\caption{The vertex characteristics of the digraph $\vec H_1(u_3)$}
\label{tb11}
\begin{tabular}{|l|l|l|}
\hline
$I_{u_1}=(1,2)$; & $I_{u_2}=(1,2)$; & $I_{u_3}=(0)$;\\
$I^{\prime}_{u_1}=(0,1)$; & $I^{\prime}_{u_2}=(0,1)$; & $I^{\prime}_{u_3}=\oslash$;\\
$O_{u_1}=(2)$; & $O_{u_2}=(2)$; & $O_{u_3}=(2,2)$;\\
$O^{\prime}_{u_1}=(1)$; & $O^{\prime}_{u_2}=(1)$; & $O^{\prime}_{u_3}=(1,1)$.\\
\hline
\end{tabular}
\end{table}
\medskip

In this case, we see that the digraphs $\vec G_1(v_2)$ and $\vec H_1(u_3)$ are positionally equivalent. The 
characteristics of pairs of vertices $(v_2,u_3)$, $(v_4,u_1)$ and $(v_5,u_2)$ are equal, and the pair of 
vertices $(v_2,u_3)$ belongs to the bijective mapping $\varphi$ between the vertices in the given graph $G$ and $H$.

\begin{theorem}
\label{thm5}
Time to compare the vertex characteristics of auxiliary digraphs, using the  history of the graph changes, 
is $O(n^3)$.
\end{theorem}

{\bf Proof}. For comparison of the vertex characteristics of the auxiliary digraphs, consisting of the single 
``floor'', it required, obviously, $O(n^2)$ time units. Therefore, for comparison of the vertex characteristics 
of the auxiliary digraphs, using the history of the graph changes, which consists of $O(n)$ ``floors'', it 
required $O(n^3)$ time.$\diamondsuit$

\section{The search algorithm}

We describe now the algorithm in more detail.

\medskip

The input of the algorithm: graphs $G=(V_G,E_G)$, $H=(V_H,E_H)\in L_n$,  
isomorphism of which it is necessary to determine if it exists. We assume that these graphs have the same number of vertices and edges, as well as their vectors 
of local degrees $D_G$ and $D_H$ are equal.

The output of the algorithm: the determining the one-to-one correspondence  
$P$ between the vertex sets of $V_G$ and $V_H$, if it exists.

\begin{list}{}{
\setlength{\topsep}{2mm}
\setlength{\itemsep}{0mm}
\setlength{\parsep}{1mm}
}
\item
{\bf The algorithm for determining the bijective mapping between the vertices in the isomorphic graphs.}

\item[{\it Step 1.}] Put $Q=G$, $S=H$, $P=\oslash$, $N:=n$, $i:=1$, $j:=1$.

\item[{\it Step 2.}] Choose the vertex $v_i\in V_Q$ in the graph $Q$. 

\item[{\it Step 3.}] Construct the auxiliary digraph $\vec Q(v_i)$, using the graph $Q$.

\item[{\it Step 4.}] Find vertex characteristics of the auxiliary digraph  $\vec Q(v_i)$.

\item[{\it Step 5.}] Choose the vertex $u_j\in V_S$ in the graph $S$.

\item[{\it Step 6.}] Construct the auxiliary digraph $\vec S(u_j)$, using the graph $S$.

\item[{\it Step 7.}] Find vertex characteristics of the auxiliary digraph $\vec S(u_j)$.

\item[{\it Step 8.}] Compare the vertex characteristics of the digraphs $\vec Q(v_i)$ and $\vec S(u_j)$ 
in the neighborhood of the vertices $v_i$ and $u_j$ of the same level.

\item[{\it Step 9}] If the graphs $\vec Q(v_i)$ and $\vec S(u_j)$ are not positionally equivalent then if $j<N$ 
then $j:=j+1$ and go to Step 5 else stop the computations, as the graphs $G$ and $H$ are not isomorphic.

\item[{\it Step 10.}] If the digraphs $\vec Q(v_i)$ and $\vec S(u_j)$ are  positionally equivalent then find 
the vertex sets $\{v_{i_1},\ldots,v_{i_t}\}$, $\{u_{j_1},\ldots,u_{j_t}\}$, unique in each of the digraphs.

\item[{\it Step 11.}] Put $P:=P\cup \{(v_{i_1},u_{j_1}),\ldots,(v_{i_t},u_{j_t})\}$, 
$V_Q:=V_Q\setminus \{v_{i_1},\ldots,v_{i_t}\}$, $V_S:=V_S\setminus \{u_{j_1},\ldots,u_{j_t}\}$, 
$N:=N-t$. 

\item[{\it Step 12.}] If $N\not= 0$, put $i:=i$, $j:=1$ and go to Step 2. Otherwise, stop the computations, 
because the bijective mapping between the vertices in isomorphic graphs $G$ and $H$ has constructed, the 
pairs of the respective vertices are stored in the set $P$.

\end{list}

\begin{theorem}
\label{thm4}
The algorithm for determining the bijective mapping the vertices in isomorphic graphs finds the mapping if it exists.
\end{theorem}

{\bf Proof}. By Theorem \ref{thm2} if graphs $G$ and $H$ are isomorphic then any 
bijective mapping $\varphi$, which convert the graph $G$ into the graph $H$ (and conversely), is 
determined by vertex pairs of the auxiliary positionally equivalent digraphs $\vec G(v)$ and $\vec H(u)$, 
having equal characteristics. If it is found $t$ such vertices, then they explicitly define the elements 
of the bijective mapping $\varphi$. Note that among the vertices of the auxiliary positionally equivalent 
digraphs at least one vertex is unique in each of the digraphs. These are vertices, which induce the 
digraphs $\vec G(v)$ and $\vec H(u)$.

Deletion of the unique vertices from the graphs $G$ and $H$ reduces to obtaining the graphs which are also 
isomorphic. Therefore, the repetition of the above procedure to the obtained graphs will lead to the 
exhaustion of the vertex list of isomorphic graphs $G$ and $H$.$\diamondsuit$

\begin{theorem}
\label{tm6}
The running time of the algorithm for determining the bijective mapping the vertices in isomorphic graphs is $O(n^5)$.
\end{theorem}

{\bf Proof}. We define the running time of the algorithm when performing steps 5--9.

Steps 5, 9 require to expend one unit of time for each step. Steps 6-7 require to expend $O(n^2)$ 
time units each. Step 8 requires to perform $O(n^3)$ time units. Therefore, $n$-multiple executing steps 
5--11 require to expend $O(n^4)$ time units.

Single execution of Steps 2--12 requires, obviously, $O(n^4)$ time units and $n$-multiple execution require 
$O(n^5)$ time units.$\diamondsuit$

\section{Conclusion}

The results, presented in this article, show the fruitfulness of the method of positioning vertices of the 
given graphs.

One can allocate two features of our method.

\begin{enumerate}
\item Each of the given graphs $G,H\in L_n$ is represented in the single form --- in the form of the auxiliary 
digraph.
\item The vertices of every digraph are positioned relative to each other without 
become attached to the vertex names of graphs.
\end{enumerate}

It is possible that this approach will be used to solve other problems.

\input{g-app}

\end{document}

%% file: g-app.tex
\newpage
{\bf \appendixname}
\vspace{1pc}
\setcounter{section}{7}
\setcounter{figure}{0}
\setcounter{table}{0}

We illustrate the proposed algorithm on another example.

\begin{figure}[ht]
\centering
\mbox{\input{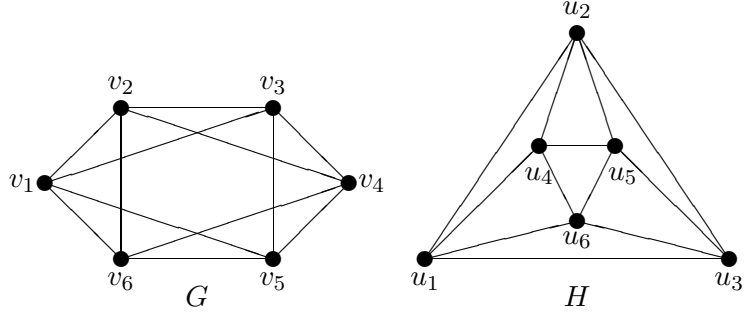}}
\caption{The given graphs $G$ and $H$.}
\label{g-a1}
\end{figure}

Let two graphs $G$, $H\in L_n$ are given, the isomorphism of which we want to determine (see Fig. \ref{g-a1}).

\begin{figure}[ht]
\centering
\mbox{\input{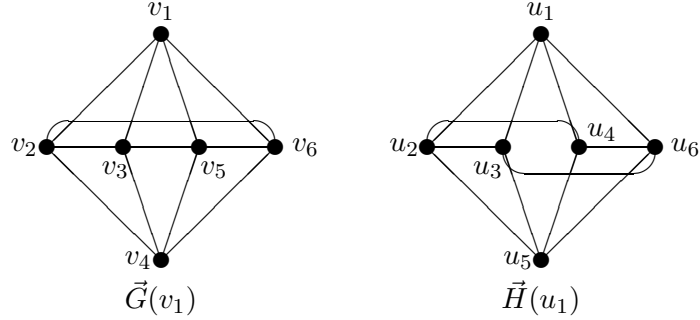}}
\caption{The auxiliary digraphs $\vec G(v_1)$ and $\vec H(u_1)$}
\label{g-a2} 
\end{figure}

Choose the vertex $v_1$ in the graph $G$ and construct the auxiliary digraph $\vec G(v_1)$ (see Fig. \ref{g-a2}).

Find vertex characteristics of the constructed digraph $\vec G(v_1)$. The results are located to the 
table \ref{tb12}.

\medskip

\begin{table}[ht]
\centering
\caption{The vertex characteristics of digraph $\vec G(v_1)$}
\label{tb12}
\begin{tabular}{|l|l|l|}
\hline
$I_{v_1}=\oslash$; & $I_{v_2}=(0,1,1)$; & $I_{v_3}=(0,1,1)$;\\ 
$O_{v_1}=(1,1,1,1)$; & $O_{v_2}=(1,1,2)$; & $O_{v_3}=(1,1,2)$;\\
\hline
$I_{v_4}=(1,1,1,1)$; & $I_{v_5}=(0,1,1)$; & $I_{v_6}=(0,1,1)$;\\
$O_{v_4}=\oslash$; & $O_{v_5}=(1,1,2)$; & $O_{v_6}=(1,1,2)$;\\
\hline
\end{tabular}
\end{table}
\medskip

Choose the vertex $u_1$ in the graph $H$ and construct the auxiliary digraph $\vec H(u_1)$ (see Fig. \ref{g-a2}).

Calculate characteristics of vertices of the newly constructed digraph $\vec H(u_1)$. The results are located to 
the table \ref{tb13}.

\medskip

\begin{table}[ht]
\centering
\caption{The vertex characteristics of the digraph $\vec H(u_1)$}
\label{tb13}
\begin{tabular}{|l|l|l|}
\hline
$I_{u_1}=\oslash$; & $I_{u_2}=(0,1,1)$; & $I_{u_3}=(0,1,1)$;\\ 
$O_{u_1}=(1,1,1,1)$; & $O_{u_2}=(1,1,2)$; & $O_{u_3}=(1,1,2)$;\\
\hline
$I_{u_4}=(0,1,1)$; & $I_{u_5}=(1,1,1,1)$; & $I_{u_6}=(0,1,1)$;\\
$O_{u_4}=(1,1,2)$; & $O_{u_5}=\oslash$; & $O_{u_6}=(1,1,2)$;\\
\hline
\end{tabular}
\end{table}
\medskip

It is easy to see that the constructed auxiliary digraphs $\vec G(v_1)$ and 
$\vec H(u_1)$ are positionally equivalent. The digraph $\vec G(v_1)$ has two unique vertices 
$v_1$ and $v_4$. They correspond to the unique vertices $u_1$, $u_5$ in the digraph $\vec H(u_1)$. 
Therefore, vertex pairs $(v_1,u_1)$, $(v_4,u_5)$ can be saved in the set $P$.

\begin{figure}[ht]
\centering
\mbox{\input{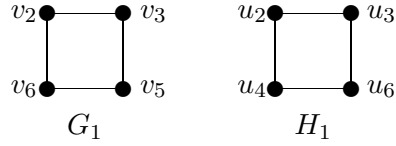}}
\caption{Graphs $G_1$ and $H_1$.}
\label{g-a3}
\end{figure}

Delete from the graphs $G$ and $H$ the unique vertices. We get graphs $G_1$ and $H_1$ (see Fig. \ref{g-a3}).

Choose the vertex $v_2$ in the graph $G_1$ and construct the auxiliary digraph $\vec G_1(v_2)$ (see Fig. \ref{g-a4}).

\begin{figure}[ht]
\centering
\mbox{\input{g-a4.pic}}
\caption{The digraphs $\vec G_1(v_2)$ and $\vec H_1(u_2)$}
\label{g-a4}
\end{figure}

Find the vertex characteristics of the constructed digraph. The results are located to the table \ref{tb14}.

\medskip

\begin{table}[ht]
\centering
\caption{The vertex characteristics of the digraph $\vec G_1(v_2)$}
\label{tb14}
\begin{tabular}{|l|l|l|l|}
\hline
$I_{v_2}=(0,1,1)$; & $I_{v_3}=(0,1,1)$; & $I_{v_5}=(0,1,1)$; & $I_{v_6}=(0,1,1)$;\\ 
$I^{\prime}_{_2}=\oslash$; & $I^{\prime}_{v_3}=(0)$; & $I^{\prime}_{v_5}=(1,1)$; & $I^{\prime}_{v_6}=(0)$;\\ 
$O_{v_2}=(1,1,2)$; & $O_{v_3}=(1,1,2)$;& $O_{v_5}=(1,1,2)$; & $O_{v_6}=(1,1,2)$;\\
$O^{\prime}_{v_2}=(1,1)$; & $O^{\prime}_{v_3}=(2)$; & $O^{\prime}_{v_5}=\oslash$; & $O^{\prime}_{v_6}=(2)$;\\
\hline
\end{tabular}
\end{table}
\medskip

Choose the vertex $u_2$ in the graph $H_1$ and construct the auxiliary digraph $\vec H_1(u_2)$ (see Fig. \ref{g-a4}).

The calculated characteristics of vertices of the newly constructed digraph are located to the table \ref{tb15}.

\medskip

\begin{table}[ht]
\centering
\caption{The vertex characteristics of the digraph $\vec H_1(u_2)$}
\label{tb15}
\begin{tabular}{|l|l|l|l|}
\hline
$I_{u_2}=(0,1,1)$; & $I_{u_3}=(0,1,1)$; & $I_{u_4}=(0,1,1)$;  & $I_{u_6}=(0,1,1)$;\\
$I^{\prime}_{u_2}=\oslash$; & $I^{\prime}_{u_3}=(0)$; & $I^{\prime}_{u_4}=(0)$; & $I^{\prime}_{u_6}=(1,1)$;\\ 
$O_{u_2}=(1,1,2)$; & $O_{u_3}=(1,1,2)$; & $O_{u_4}=(1,1,2)$; & $O_{u_6}=(1,1,2)$;\\
$O^{\prime}_{u_2}=(1,1)$; & $O^{\prime}_{u_3}=(2)$; & $O^{\prime}_{u_4}=(2)$; & $O^{\prime}_{u_6}=\oslash$;\\
\hline
\end{tabular}
\end{table}
\medskip

We find that the constructed auxiliary digraphs $\vec G_1(v_2)$ and $\vec H_1(u_2)$ are positionally equivalent. The digraph $\vec G_1(v_2)$ has two unique vertices $v_2$ and $v_5$. They correspond to the unique vertices $u_2$, $u_6$ in the digraph $\vec H_1(u_2)$. Therefore, vertex pairs $(v_2,u_2)$, $(v_5,u_6)$ can be saved in the set $P$.

Removing the unique vertices of the graphs $G_1$ and $H_1$, we obtain respectively the graphs $G_2$ and $H_2$ consisting of two isolated vertices each. It is clear that it will be obtained auxiliary positionally equivalent  digraphs $\vec G_2(v_3)$ and $\vec H_2(u_3)$. Inducing vertices form the pair $(v_3,u_3)$ of unique vertices that can be saved in the set $P$.

We also find that the pair of vertices $(v_6,u_4)$ belongs to the set $P$.

Thus, the binary relation between vertices of isomorphic graphs $G$ and $H$ is:
\[
\varphi=\{(v_1,u_1),(v_2,u_2),(v_3,u_3,),(v_4,u_5),(v_5,u_6),(v_6,u_4)\}.
\]

Perform the verification of the result.
\medskip

\begin{center}
\begin{tabular}{cc}
\begin{tabular}{c}
$\{v_1,v_2\} \leftrightarrow \{u_1,u_2\}$ \\
$\{v_1,v_3\} \leftrightarrow \{u_1,u_3\}$ \\
$\{v_1,v_5\} \leftrightarrow \{u_1,u_6\}$ \\
$\{v_1,v_6\} \leftrightarrow \{u_1,u_4\}$ \\
$\{v_2,v_3\} \leftrightarrow \{u_2,u_3\}$ \\
$\{v_2,v_4\} \leftrightarrow \{u_2,u_5\}$ \\
\end{tabular}
&
\begin{tabular}{c}
$\{v_2,v_6\} \leftrightarrow \{u_2,u_4\}$ \\
$\{v_3,v_4\} \leftrightarrow \{u_3,u_5\}$ \\
$\{v_3,v_5\} \leftrightarrow \{u_3,u_6\}$ \\
$\{v_4,v_5\} \leftrightarrow \{u_5,u_6\}$ \\
$\{v_4,v_6\} \leftrightarrow \{u_5,u_4\}$ \\
$\{v_5,v_6\} \leftrightarrow \{u_6,u_4\}$ \\
\end{tabular}\\
\end{tabular}
\end{center}
\medskip

All the matching edges of the source graphs are found.